\documentclass[%
 reprint,
 amsmath,amssymb,
 aps,
]{revtex4-2}

\usepackage{graphicx}
\usepackage{dcolumn}
\usepackage{bm}

\begin{document}

\preprint{APS/123-QED}

\title{Laboratory Demonstration of Decentralized, Physics-Driven Learning}

\author{Sam Dillavou}
\author{Menachem Stern} 
\author{Andrea J. Liu}
\author{Douglas J. Durian}
 \affiliation{Department of Physics and Astronomy, University of Pennsylvania, 209 South 33rd Street, Philadelphia, PA 19104, USA}

\date{\today}

\begin{abstract}
In typical artificial neural networks, neurons adjust according to global calculations of a central processor, but in the brain neurons and synapses self-adjust based on local information. Contrastive learning algorithms have recently been proposed to train physical systems, such as fluidic, mechanical, or electrical networks, to perform machine learning tasks from local evolution rules. However, to date such systems have only been implemented {\it in silico} due to the engineering challenge of creating elements that autonomously evolve based on their own response to two sets of global boundary conditions. Here we introduce and implement a physics-driven contrastive learning scheme for a network of variable resistors, using circuitry to locally compare the response of {\it two} identical networks subjected to the two different sets of boundary conditions. Using this innovation, our system effectively trains itself, optimizing its resistance values without use of a central processor or external information storage. Once the system is trained for a specified allostery, regression, or classification task, the task is subsequently performed rapidly and automatically by the physical imperative to minimize power dissipation in response to the given voltage inputs. We demonstrate that, unlike typical computers, such learning systems are robust to extreme damage (and thus manufacturing defects) due to their decentralized learning. Our twin-network approach is therefore readily scalable to extremely large or nonlinear networks where its distributed nature will be an enormous advantage; a laboratory network of only 500 edges will already outpace its {\it in silico} counterpart. 
\end{abstract}

\maketitle


The confluence of ideas from neuroscience and machine learning has contributed immensely to our fundamental understanding of the nature of learning~\cite{richards_deep_2019,hasson_direct_2020}. However, biological neural networks differ fundamentally from standard machine learning algorithms in an important way~\cite{tavanaei_bpstdp_2019,ganguly_energy_2019}. 
A typical artificial neural network (ANN) requires a processing unit (\textit{e.g.} CPU) that trains the network by minimizing a global cost function~\cite{lecun_deep_2015}, while repeatedly storing and retrieving information from a separate electronic memory. This von~Neumann architecture is very successful but creates a severe computational bottleneck. In contrast, the brain and other biological networks~\cite{tero_physarum_2006, alim_mechanism_2017} are more akin to extremely sophisticated and adaptive metamaterials: they are physical systems made of repeated, locally responsive elements (\textit{e.g.} neurons and synapses) that generate learning as a highly complex emergent property. This distribution and parallelization of computation and memory storage allows the human brain ($\sim10^{11}$ neurons and $\sim10^{14}$ synapses) to function at reasonable speeds despite signal propagation timescales millions of times slower than modern computational clock cycles. Furthermore, it allows the brain to recover from massive damage~\cite{mcgovern_hemispherectomy_2019} while consuming only modest power~\cite{sengupta_power_2014} compared to typical computers. 

These advantages of the brain have spurred efforts to imitate its features~\cite{bengio_biologically_2015, bengio_early_2016, bengio_stdp_2015,markovic_physics_2020}. Several of these have only been realized \emph{in silico}~\cite{ernoult_using_2019,kendall_training_2020, martin_eqspike_2021} or in hybrid \emph{in situ}-\emph{in silico} form~\cite{li_efficient_2018, yao_fully_2020, wright_deep_2022}.  Actual laboratory realizations of `neuromorphic' hardware that bypass processors tend to mimic either standard machine learning algorithms (\textit{e.g.} back-propagation)~\cite{hu_dotproduct_2016, wang_reinforcement_2019, zhang_neuroinspired_2020} or phenomenological synaptic rules found in the brain (\textit{e.g.} spike-timing-dependent plasticity)~\cite{arima_refreshable_1992,schneider_analog_1993, kim_experimental_2015, labarbera_interplay_2016, serb_unsupervised_2016}. 

An alternate approach to learning without a processor is to exploit \emph{physical processes} in tandem with simple \textit{and} local rules. Laboratory mechanical networks have been trained without any sort of processor to develop negative Poisson's ratios using the process of `directed aging'~\cite{pashine_directed_2019, hexner_effect_2020}, which exploits the natural physical tendency of a mechanical network to minimize elastic energy when a stress is applied.  `Contrastive learning'~\cite{movellan_contrastive_1991} compares the response of the system to two different boundary conditions to adjust the degrees of freedom; this works more robustly than directed aging in laboratory mechanical networks~\cite{pashine_local_2021}, but has thus far required an external entity to enact these local rules. The ‘equilibrium propagation’ framework ~\cite{scellier_equilibrium_2017,kendall_training_2020,scellier_deep_2021b} can be viewed as combining the concept of directed aging with contrastive learning and specifies simple local learning rules that in principle can be implemented in flow networks ~\cite{kendall_training_2020}. Equilibrium propagation nudges the network towards the desired target solution instead of imposing it directly; in the limit of infinitesimal nudges, the learning rule performs gradient descent on a loss function. A framework known as ‘coupled learning’ ~\cite{stern_supervised_2021} builds on equilibrium propagation, providing the foundation for our work. In both frameworks, although the learning rules are spatially local, they require simultaneous access to two distinct states of the same system. As a result, they are not \emph{temporally} local, and require the use of memory when implemented \textit{in silico}. This issue has thus far prevented them from being realized in the laboratory.

\begin{figure}[h!b] 
\includegraphics[width = 8.7cm]{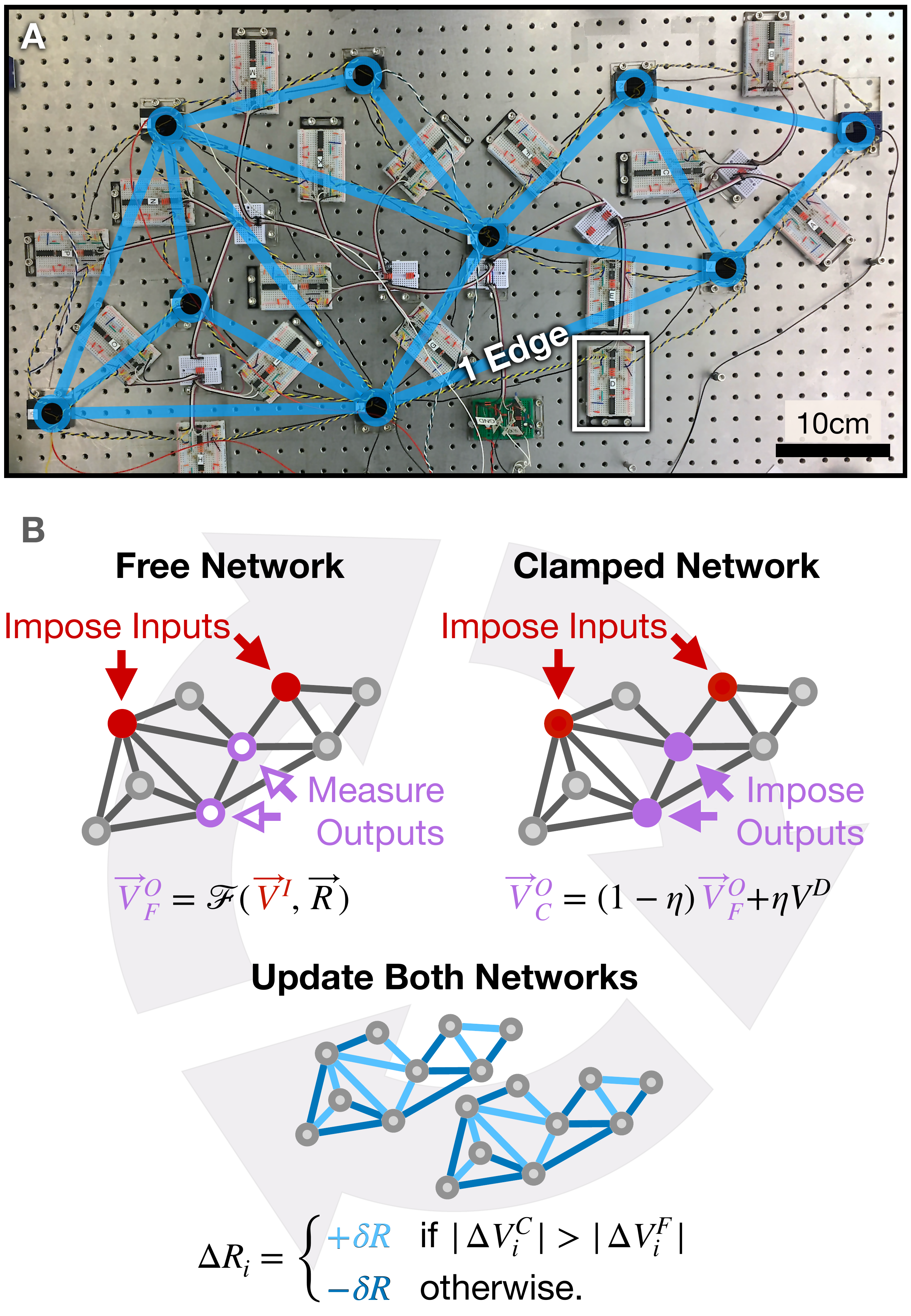}
\caption{\textbf{A Physics-Driven Learning Machine} 
\textbf{(A)} Image of the 16-edge circuitry, with the network structure overlaid in blue. Each breadboard, like the one highlighted in white, houses commensurate edges in the free and the clamped network. For circuitry details see Appendix C.
\textbf{(B)} Procedure for training the learning machine. A supervisor (i) imposes voltages to the inputs (red) in the free network and (ii) to the inputs and outputs (purple) in the clamped network. The network (iii) updates its own resistances, and $\vec V^O_F$ is `calculated' by physical laws.
}
\label{fig1}
\end{figure} 

In this study we report the laboratory realization of a physical learning machine composed of a pair of variable resistor networks. We resolve the highly restrictive and challenging requirement of contrastive learning in physical systems by using two identical twin networks to simultaneously measure responses of the `same' physical system to two different sets of boundary conditions. When we expose the system to training data, the physical imperative to minimize energy dissipation carries out the forward calculation to `compute' the outputs within nanoseconds, while local rules that adjust the resistances of the edges take the place of backpropagation, obviating the need for a processor or memory storage. We demonstrate that such a network can learn to perform and switch among a variety of tasks, including allostery, regression, and classification. Finally, we show that because the learning is fully distributed and each edge learns individually, the network functionality is highly robust to network changes and damage, making it readily scalable.

\section*{Approach}
In previous work, simulated and laboratory mechanical networks, and simulated flow networks, have been trained to perform desired tasks by adjusting their internal degrees of freedom~\cite{goodrich_principle_2015,rocks_designing_2017,stern_shaping_2018,rocks_limits_2019,ruiz-garcia_tuning_2019,scellier_equilibrium_2017,pashine_directed_2019,hexner_periodic_2020,hexner_effect_2020,stern_continual_2020,stern_supervised_2020,kendall_training_2020,stern_supervised_2021,pashine_local_2021}. This has been accomplished either by minimizing a global cost function~\cite{goodrich_principle_2015,rocks_designing_2017,stern_shaping_2018,rocks_limits_2019,ruiz-garcia_tuning_2019} or using local rules aided by an external processor~\cite{scellier_equilibrium_2017,pashine_directed_2019,hexner_periodic_2020,hexner_effect_2020,stern_continual_2020,stern_supervised_2020,kendall_training_2020,stern_supervised_2021,pashine_local_2021}. Here we
consider a \textit{self-adjusting} electronic network comprised of nodes connected by variable resistors, whose values we will call the ``learning degrees of freedom." When voltages $\vec V^I$ are applied at input nodes, the voltages at designated output nodes $\vec V^O$ are physically determined as functions of the input voltages and the resistance values $\vec R$ of the network edges, as the system minimizes the total energy dissipation. The coupled learning~\cite{stern_supervised_2021} framework for supervised learning specifies local evolution rules for how each resistance should evolve to produce desired output voltages. In doing so, the system exploits the physical processes that govern the network to perform computation, and implements contrastive learning as a spatially local rule, in a similar manner as equilibrium propagation~\cite{scellier_equilibrium_2017}.

In supervised learning, training examples determine the inputs $\vec V^I$ as well as the \emph{desired} output responses $\vec V^D$ for each example. These desired output voltages can be achieved by adjusting the resistances of all the edges, $\vec R$ (the \textit{learning degrees of freedom}). During training $\vec R$ are adjusted based on a comparison of two distinct electrical states imposed on the same network. In the \textit{free} state, the network attempts the desired task: input voltages $\vec V^I$ are applied, and the network produces output voltages $\vec V^O_F$. In the \textit{clamped} state, the same inputs $\vec V^I$ are applied, but voltages are also applied at the output nodes; those voltages are \textit{clamped} at values $\vec V^O_C$ closer to the desired values than $\vec V^O_F$:
\begin{equation}
    \vec V_C^O = \eta \vec V^D + (1-\eta) \vec V_F^O
    \label{eta}
\end{equation}
where $0<\eta\leq 1$ is the amplitude of the nudge toward the desired state. 

When input voltage values $\vec V^I$ are applied to the network, physical laws adjust all other node voltages--which we call the ``physical degrees of freedom," to minimize total energy dissipation (the ``physical cost function"). Therefore clamping the outputs nodes away from their ``free" state towards the goal both requires additional power and creates a lower error electrical state. Small adjustments to the learning degrees of freedom, $\vec R$, that lower the energy dissipation of the (higher-power) clamped state $P^C$ relative to the (lower-power) free state $P^F$ will create a new free state equilibrium with output voltages that lie between the old free and clamped states. Determining the direction to update each resistance requires only spatially local information, namely which state (free or clamped) has a higher energy dissipation (voltage drop) across the edge in question, allowing edges to update their own resistance. Similar to equilibrium propagation, this algorithm approximates global gradient descent in the limit $\eta \ll 1$~\cite{stern_supervised_2021}, allowing a system to \textit{train itself} by repeating this update process. However, this algorithm is not \textit{temporally} local, in that it requires simultaneous access to the 
response for two distinct sets of boundary conditions which, by definition, cannot be imposed simultaneously. It is this requirement that makes contrastive learning in physical systems so challenging to realize.

Here we resolve this conundrum by building two identical electrical networks to run the free and clamped states. We use digital variable resistors (see Methods) on each edge, which have 128 possible discrete resistance values. The original (continuous) coupled learning update rule,
\begin{equation}
    \Delta R_i = \frac{\gamma}{R_i^2}\big([\Delta V^C_i]^2 - [\Delta V^F_i]^2\big)
    \label{originalcouple}
\end{equation}
where $\gamma$ is a learning rate and $\Delta V^C_i$, $\Delta V^F_i$ are the voltage drops in edge $i$ of the clamped and free states respectively. In our discrete resistor networks, the two networks adjust their (identical) resistances according to an approximation of the original rule,
\begin{equation}
    \Delta R_i^C =   \Delta R_i^F = 
     \begin{cases}
      +\delta R & \text{if} \ |\Delta V^C_i| > |\Delta V^F_i|, \\
      -\delta R & \text{otherwise.}
    \end{cases}  
    \label{sign}
\end{equation}
equivalent to taking the sign of Eq.~(\ref{originalcouple}) multiplied by $\gamma = \delta R$. This now Boolean operation is carried out by integrated circuits housed on each edge of the network; the entire system is pictured in Fig.~\ref{fig1}A. For details regarding the implementation of this rule, see Appendix~C. Because the learning process is decentralized, our system functions without a central processor, and training the network to perform a task is straightforward. The procedure is detailed in Fig.~\ref{fig1}B: apply the desired input voltages to the free and clamped networks, as well as clamped output voltages $\vec V^O_C$ to the clamped network. Edge updates are triggered by a global clock, and no further instruction to the edges are required, as each edge is responsible for its own evolution.

\begin{figure} 
\includegraphics[width = 8.7cm]{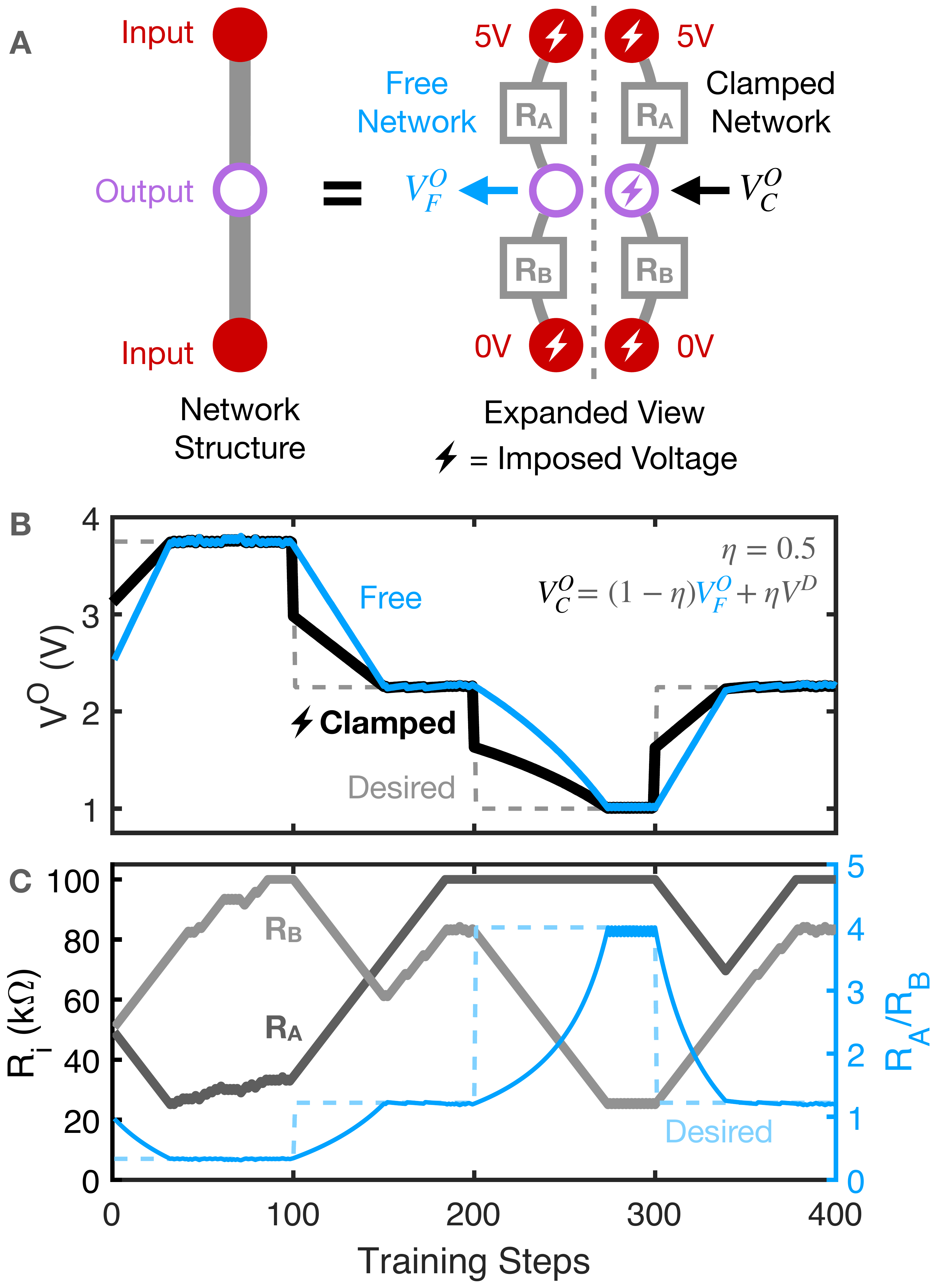}
\caption{\textbf{A self-training voltage divider.} \textbf{(A)} Diagram of network structure, as depicted in later figures (left) and expanded (right) to show both free and clamped networks. Voltage is imposed on input nodes (red) in both networks, and on output nodes (purple) only in the clamped network. The resistance of each edge is identical in both networks. \textbf{(B)} Output node voltage $V^O$ vs training steps for both free (blue) and clamped (black) networks. The desired voltage $V^D$ is shown as a gray dashed line. Note that the clamped state effectively guides the free state towards the desired voltage which is changed every 100 steps, from 3.75~V, to 2.25~V, to 1~V, and finally to 2.25~V. \textbf{(C)} Resistance values of the two edges in the network (grays) and their ratio (blue) as a function of training step. The light blue dashed line represents the ratio that will produce the desired network output.}
\label{figS2}
\end{figure} 

To demonstrate operation of our learning elements, we train a two-edge network (Fig.~\ref{figS2}A) as a voltage divider: We ask the network to produce a single desired voltage $V^D$ at its output (middle) node, while the input nodes (top and bottom) are held at 5~V and 0~V respectively. To train, the following algorithm is repeated every clock cycle:
\begin{enumerate}
  \item Update the clamped state output node voltage, per Eq.~(\ref{eta}).
  \item Every edge updates its own resistance, per Eq.~(\ref{sign}).
\end{enumerate}
In machine learning language, the `supervisor' tells the network the right answer through the clamped boundary condition. The network itself decides \textit{how} to achieve this answer, as it receives no external instructions about which edges to push up or down in resistance. That is, shown the right answer, the network \textit{trains itself} to produce it. In this simple example, this distinction may seem trivial, but as we increase the size of the network, the job of the supervisor does not grow in complexity; it is always given by Eq.~(\ref{eta}). This is in stark contrast to ANNs, where the number of gradient calculations grows rapidly with network size.

As previously described, edges modify their resistance to bias the electrical state of the system away from the free state and towards the clamped state. This results in the free state output voltage(s) `following' the clamped state voltages, which in turn move progressively towards the desired voltage (Fig.~\ref{figS2}B). In our voltage divider, the desired voltage was changed every 100 training steps.  At the start, all edges are initialized at the center of their resistance ranges ($\sim 50~{\rm k}\Omega$).  Two phases in each training are evident. At first, the clamped and free networks are quite different, and the two edges evolve in opposite directions until the desired voltage is achieved (Fig.~\ref{figS2}C). Once the network has reduced the error sufficiently, noise dominates the signal to the comparators, resulting in occasional incorrect evaluations when comparing voltages differing by less than 0.01~V (as mentioned previously, and shown in Appendix~C, Fig.~5D). These occasional errors create an error floor, but also allow the network to explore the phase space of valid solutions; the ratio of the two resistance values (blue line in Fig.~\ref{figS2}C) remains nearly constant while both resistance values drift. For more complex networks and tasks this stochasticity may be useful; similar exploration of the available solutions space can promote generalization in both biological~\cite{kappel_network_2015} and artificial networks~\cite{feng_inverse_2021}.

\begin{figure} 

\includegraphics[width = 8.5cm]{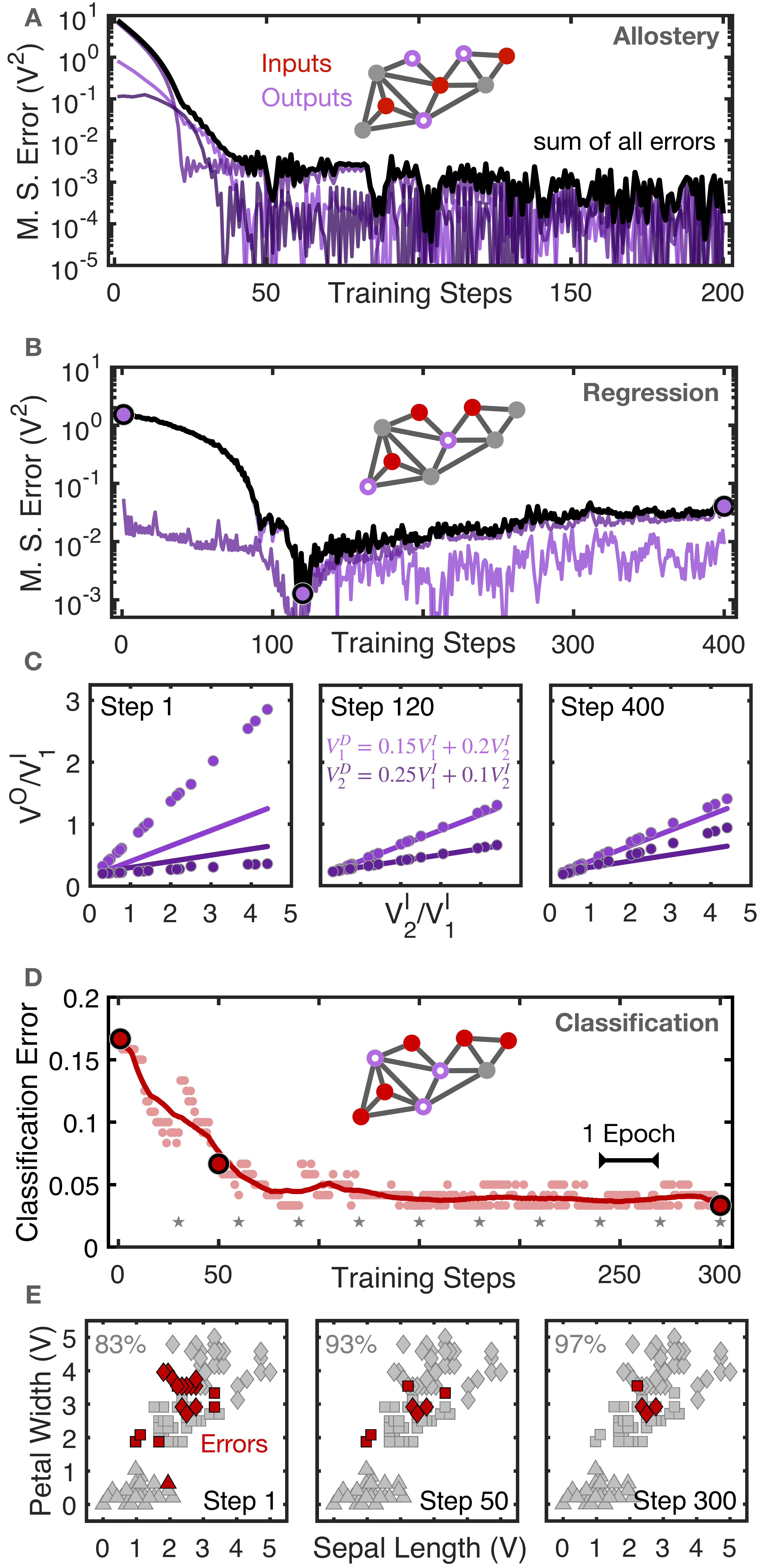}
\caption{\textbf{One physical system performs many tasks.} 
\textbf{(A)} Mean-squared error for each of three outputs and their sum (black) vs training step for an example allostery task. 
\textbf{(B)} Mean-squared error for each of two outputs and their sum (black) for a two-parameter regression task for each output node. Large purple circles indicate the training steps shown in \textbf{C}. 
\textbf{(C)} Snapshots of the values for both outputs at three steps during training for the regression task in \textbf{B}. Lines indicate the desired output values. Regression involves two parameters, and thus both axes are scaled by $V^I_1$ to project the results into 2 dimensions. 
\textbf{(D)} Test set classification error for the iris benchmark dataset~\cite{fisher_use_1936} vs training step (faded symbols). Smoothing the data with a window of 30 training steps (solid line) highlights that the final plateau accuracy is above 95\%. Large red circles indicate the training steps shown in \textbf{E}. The desired voltage for each class is re-measured every epoch, indicated by the gray stars (see Appendix A for details). 
\textbf{(E)} Snapshots of the classification success of the test set projected into the 2D space of two of the four inputs (sepal length and petal width, rescaled to 0-5V). Species of iris is denoted by marker shape. Gray shapes are correctly classified, red are incorrectly classified. 
}
\label{fig2}
\end{figure} 

\section*{Results}
We now demonstrate the success of our system by training a 16-edge network (Fig.~\ref{fig1}A) to perform three types of tasks inspired by biology (allostery), mathematics (regression), and computer science (classification).  Then we demonstrate its flexibility and robustness.

Allostery is a common feature of proteins~\cite{rocks_designing_2017}, in which an input signal, namely strain applied to a local region of the protein by binding a regulatory molecule, gives rise to a desired strain or conformational change elsewhere in the protein, enabling or preventing binding of a substrate molecule. In a related problem of `flow allostery'~\cite{rocks_limits_2019,rocks_revealing_2020,rocks_hidden_2021}, a pressure drop in one region of a flow network, (\textit{e.g.} across input arteries in the brain vascular network) gives rise to desired pressure drops elsewhere in the brain at designated output locations that can be quite distant from the input arteries, allowing the vascular system to deliver enhanced blood flow and therefore more oxygen to active parts of the brain. In the context of electrical networks, allostery corresponds to producing specified output voltages in response to given input voltages. This functionality can be useful for tasks such as allocating power to various connected devices.

We choose a three-input, three-output allosteric task as an example (Fig.~\ref{fig2}A inset). Using a nudge of $\eta = 0.5$, the network successfully learns to deliver 3~V at all output nodes, in response to three simultaneous input node voltages of 5, 1, and 0~V. The  mean-squared error for this task drops during the learning process by over four orders of magnitude (Fig.~\ref{fig2}A). We note that in theoretical treatment $\eta \ll 1$ is assumed; $\eta \sim 1$ will in effect be taking a finite-difference gradient with a large step size, and thus substantially degrade accuracy~\cite{stern_supervised_2021}. However, in a physical system, noise (order 0.01V) will dominate the learning process if $\eta$ is too small. Thus the success of the network at finite $\eta$ is a nontrivial demonstration of its feasibility in real systems.

Regression is a more difficult test because the desired output voltages are not constants but rather functions of the input voltages. We ask the network to solve two equations for two unknowns, choosing the two equations 
\begin{equation}
    V^D_1 = 0.15V^I_1 + 0.20 V^I_2 \quad \quad V^D_2 = 0.25V^I_1 + 0.10 V^I_2
\end{equation}
We generate a data set of 420 randomly chosen input pair values between 1 and 5~V, and calculate the desired voltage for each input pair using the above equations. We set an additional input node at 0~V to remove the freedom for a global shift in voltage, resulting in three input and two output nodes (Fig.~\ref{fig2}B inset). We divide the data into a training set (400 elements) and a test set (20 elements). Every clock cycle, the network is shown a new example from the training set, and it updates its resistance values accordingly. Between these examples, the network is given the entire test set one by one, and its free state outputs are recorded as an indication of the network's performance. Given these conditions and $\eta = 0.2$, our learning machine reduces the mean-squared error for the entire test set by over two orders of magnitude (Fig.~\ref{fig2}B), producing an accurate result despite its small size (Fig.~\ref{fig2}C). Note that during training the network finds an extremely good fit to the data around step 120, but cannot maintain it due to some combination of noise, sampling error from sequential training, and small bias in the internal logic circuitry of the edges. The observed rise in test error before the final plateau is a common feature in machine learning~\cite{yao_early_2007}.

Data classification is an even more stringent test of the network. We use a benchmark data set of three species of iris flowers~\cite{fisher_use_1936}. The network is tasked with classifying these flowers based on four measurements: petal and sepal length and width. We withhold 120 of the 150 flowers as a test set, and train on 30 flowers, 10 from each species. We designate 5 input nodes (one for each measurement plus one fixed ground) and three output nodes (Fig.~\ref{fig2}D inset). Between training steps, the entire test set of 120 flowers is run through the network, and a flower is considered correctly classified if its three outputs are closest ($L_2$ norm) to the desired outputs of the correct species. We implement a custom output scheme in which the desired outputs for a given species are recalculated every epoch by averaging outputs of that species. This provides protection against training towards infeasible outputs, and robustness to initial conditions. See Appendix A for full task specification and training details. Using this algorithm with $\eta = 0.1$, the network is able to classify the iris dataset with over 95\% accuracy (Fig.~\ref{fig2}D). For comparison, a linear classifier trained using logistic regression on this data achieves a test accuracy of 98\%.  The 2D projection of the 4D input data (Fig.~\ref{fig2}E) shows that incorrectly classified flowers lie along overlapping edges of class clusters.

\begin{figure*}[h!t] 
\includegraphics[width = 1\textwidth]{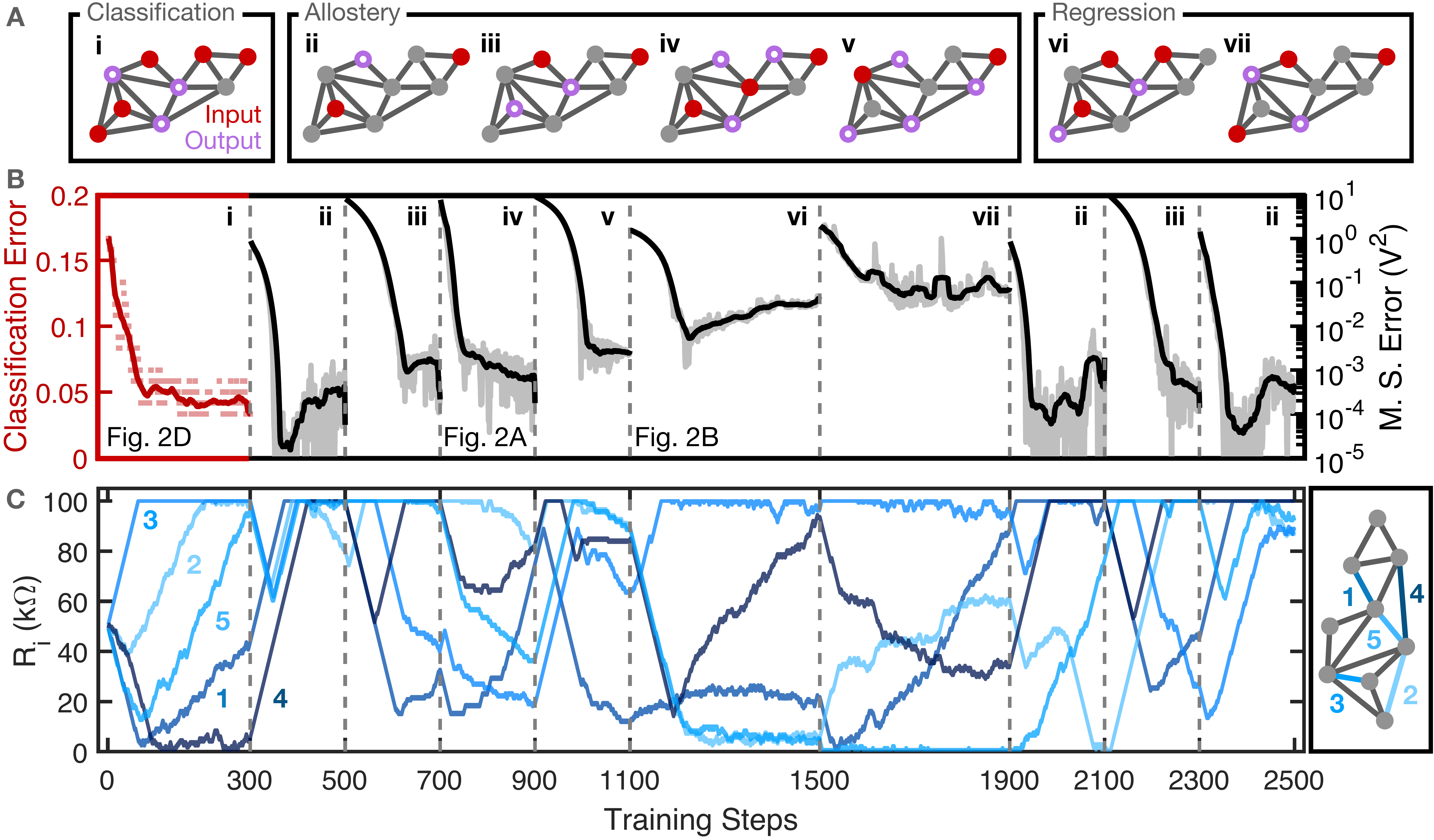}
\caption{\textbf{The learning machine is flexible and retrainable.} 
\textbf{(A)} Network structure with input (solid red) and output (purple outlined) nodes indicated for seven distinct tasks. Further details for each task are given in Appendix A. 
\textbf{(B)} Classification error (task i) and mean-squared error (tasks ii-vii) vs training step. Data is smoothed over a window of 30 training steps, with raw data shown faded in the background. The network performs tasks i-vii in order, then tasks ii, iii, and ii again. 
\textbf{(C)} Left: Resistance values of five numbered edges over the entire training process. Right: Network structure with these five numbered edges highlighted.
}
\label{fig3}
\end{figure*} 

We now highlight some features of the system. The first is the ability to learn new tasks. Unlike simulated networks, a physical learning machine must be physically manufactured. Therefore a given network is far more useful if it can switch from one task to another on demand. For our system, there is no imposed direction of information travel as in a feed-forward neural network, so any node can be used as an input node, output node, or hidden node. We demonstrate this flexibility by training our network to perform seven distinct tasks in succession, using different input-output configurations (Fig.~\ref{fig3}A). In this sequence, our 16-edge network performs one classification task (i), 4 allosteric tasks with numbers of output nodes ranging from 1 to 4~V (ii-v), and two 2-parameter linear regression tasks (vi-vii). The network successfully learns each task in turn, as indicated by the reductions in mean-squared error (Fig.~\ref{fig3}B). The edges are not reset between tasks, but simply find new values as the network adjusts to its new task and training examples (Fig.~\ref{fig3}C). Because of this ability to retrain using any input-output combination, a network does not need to be designed specifically to perform certain tasks -- it can be trained on \emph{any} task that can be framed in terms of input and output voltages. This flexibility stems, in part, from the ability of the system to `solve' a problem in multiple ways. In this sequence of tasks, our 16-edge network performs task~ii, an allosteric task with one output, three different times. Each time the solution involves different values of edge resistances $\vec R$, and furthermore explores this space of approximately equally-valid solutions that lie within the noise floor (Fig.~\ref{fig3}C). We purposefully bias this drift of resistor values to increase on average (see Appendix C), which pushes the network to avoid high-power solutions that may strain or damage hardware or waste energy. The network quickly erases memory of previous tasks, as is typical in linear networks~\cite{kirkpatrick_overcoming_2017,stern_continual_2020}, as seen by the similar initial error in performing task~ii each time (Fig.~\ref{fig3}B). The `capacity' of these networks (\textit{e.g.} the maximum number of trainable output nodes as a function of the number of nodes and edges in the system) and their ability to retain memory of previous tasks, are subjects for future work.

\begin{figure} 
\includegraphics[width = .5\textwidth]{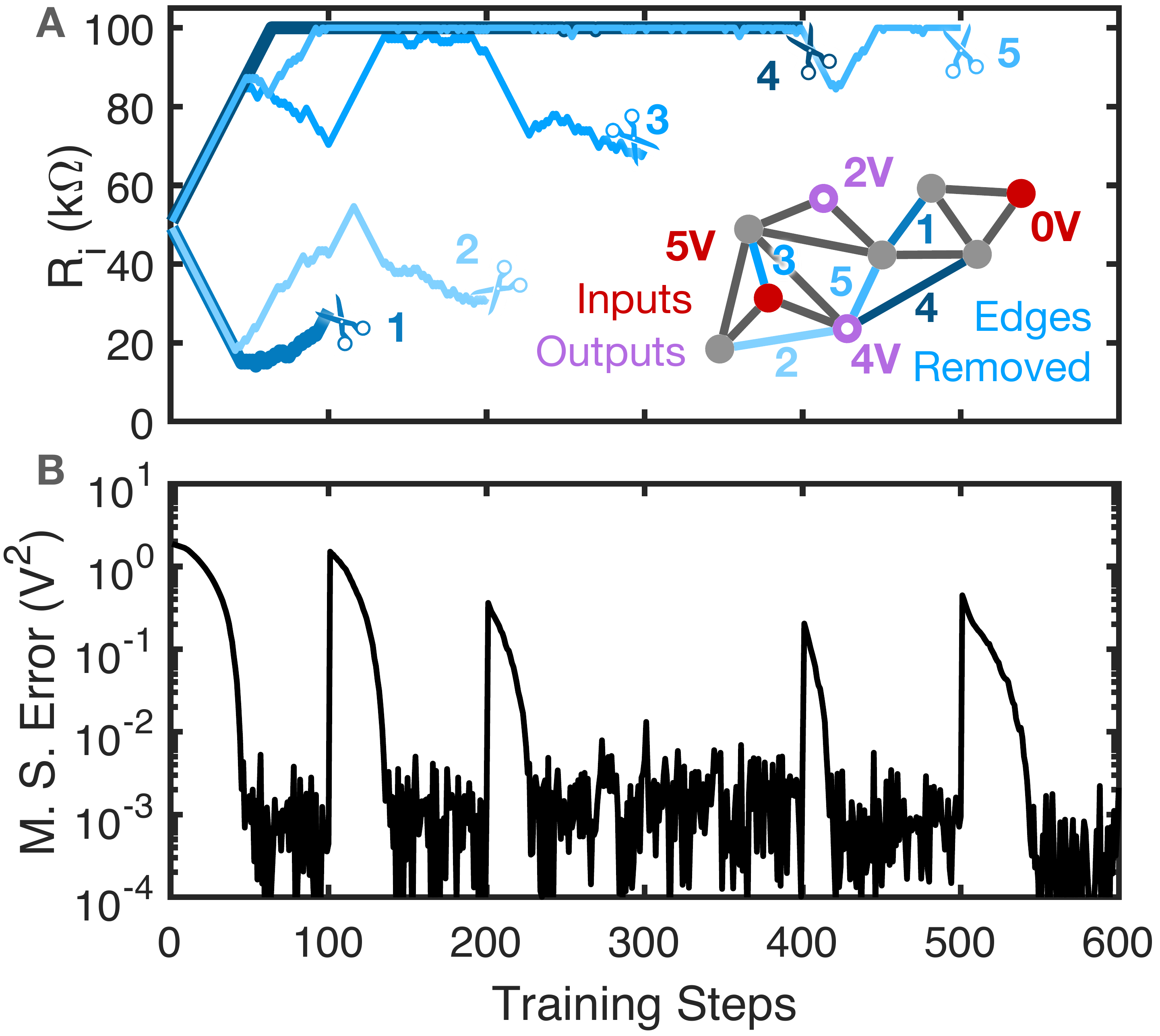}
\caption{\textbf{The learning machine is robust to damage.} 
\textbf{(A)} Resistance vs training steps for an allosteric task, as edges are cut. Inset: network structure with edges numbered by order of removal. Voltage values indicate the task being performed. Input nodes are solid red, output nodes are outlined purple. 
\textbf{(B)} Mean-squared error for this two-output allostery task vs training step. Note the spikes every 100 training steps, when an edge is removed, followed by recovery.
}
\label{fig4}
\end{figure}

A second useful feature of our network as a learning system is its robustness to damage. Physical systems used to implement simulated neural networks, such as CPUs, are usually quite fragile. Breaking or removing part of a computer usually disables it completely. In contrast, biological systems can often function despite massive damage; given the right conditions, a plucked flower not only survives, it can generate an entirely new plant. While our system cannot grow new edges, it can easily recover its desired function after substantial damage. To demonstrate this feature, we train our network to perform a 2-output allosteric task (Fig.~\ref{fig4}A inset). We track resistance values of five edges (Fig.~\ref{fig4}A), removing one every 100 training steps. During training, our 16-edge network reduces the mean-squared error of the outputs by several orders of magnitude from its initial value (Fig.~\ref{fig4}B). Removing an edge can produce an immediate spike in error as the currents adjust to the new network structure. However, the network recovers each time by finding a novel solution to the task \textit{even after nearly 1/3 of the network structure is destroyed}. Because the network is homogeneous, no edge is special, and no single part of the network is essential to its proper functioning. We note that in this demonstration, we pruned edges empirically found to be important; that is, we chose edges whose removal produced a spike in error. In fact, a substantial fraction of edges produce only a modest change in error when pruned, as is the case for the third pruned edge. Even in this linear system, memories can be robust enough against damage that retraining often isn't even needed.

\section*{Discussion}

We have built a flexible, robust, physics-driven learning circuit that learns complex tasks by adjusting its internal elements without top-down instruction from a human or computer. Even with only 16 edges, it is capable of a variety of tasks unspecified in its design, namely classification, regression, and allosteric functionality. 

Four key concepts underlie the system. First, the edge resistances are `learning' degrees of freedom, distinct from the node voltages, which are `physical' degrees of freedom. Physics constantly adjusts the physical degrees of freedom to minimize energy dissipation while the learning degrees of freedom are only adjusted during training. Then they are frozen, preserving the ability to perform the learned task. Second, there are more than enough learning degrees of freedom even in our small system to satisfy all the constraints applied in the task examples. This is why the system is able to satisfy all the tasks~\cite{rocks_limits_2019} and why it is robust to substantial damage.  Third, our approach specifies a local rule for adjusting the learning degrees of freedom that approximates minimization of the cost function~\cite{stern_supervised_2021,kendall_training_2020}. The cost functions themselves are different for different learning tasks, but the form of the learning rule, \textit{i.e.} the adjustment procedure of each edge, remains the same for any task. This is why the system can learn new tasks. Fourth, the implementation of two identical networks resolves a nontrivial constraint for contrastive learning, wherein two states of a single system, corresponding to different boundary conditions, must be compared. Our implementation does not require any additional on-board memory storage, help from a CPU, or use of temporal signals to enact. As such it is massively scalable and robust to operate. It is also robust to manufacturing errors. There is still much to be understood about even our modest 16-edge system, but the simplicity of its local rules and basis in well-understood physical laws suggest the possibility of understanding exactly what and how it learns~\cite{holzinger_machine_2018,rocks_revealing_2020,rocks_hidden_2021}.  Certainly, theoretical understanding seems less difficult to attain for the physical learning machine than for many other neuromorphic realizations, not to mention the brain itself.


Although the abilities of our current prototype are modest compared to artificial neural networks, the successful realization of a physical learning machine opens numerous paths for future work. Potentiometers with more (or continuous) states, as well as logarithmic or pseudo-logarithmic spacing of the resistance values, will greatly improve the network flexibility and reduce the error floor~\cite{stern_supervised_2021}. Diodes or other non-linear circuit elements will allow the system to perform currently-prohibited operations such as mimicking an XOR gate~\cite{minsky_perceptrons_2017,kendall_training_2020}. Importantly, we can improve both the network size and speed while diminishing the size of the components. Our largest network has only 16 edges, each on its own breadboard, and takes up several square feet. Our voltage application and measurement hardware limits the network to steps at 3--5~Hz, but the network itself is capable of operating multiple orders of magnitude faster. Furthermore, due to its Boolean logic and simultaneous comparison of two networks, as opposed to the use of memory or temporal signals, and its robustness to damage and thus manufacturing defects, the system is massively scalable. We estimate that the system can easily be scaled up in the number of edges and in the frequency of training steps by at least six orders of magnitude using readily available circuit fabrication methods~\cite{yeap_5nm_2019}. Such a circuit would have a footprint five orders of magnitude smaller than our prototype (see Appendix B for back-of-envelope calculations of these numbers). 

In computational neural networks, the computation time increases rapidly with the number of edges. An exciting feature of our system is that adding edges to the network does not increase computation time per training step, since all edges perform their own adjustments completely in parallel. This feature arises because outputs are not computed but are physical responses to stimuli, and because the job of imposing the clamping voltages does not increase in complexity as the network grows. The speed of learning depends on the physical size of the system and its inherent (tiny) capacitance, which together determine the timescale at which the voltages reach equilibrium (of order nanoseconds in our system). In the current prototype, this is far faster than our clock cycle time and thus does not affect training times. Furthermore, due to its non-specific structure, flexibility, and ability to withstand to damage, the scaling of our system is robust to imperfections and defects that invariably seep in when the number of components increase. It is possible that this ready scalability of physical learning machines may one day allow them to compete with computational neural networks. Already, with a modest increase of x100 in network size with no speed change, our prototype would outperform a simulation implementation as in \cite{stern_supervised_2021} due to the simulation's inherent bottleneck of relying on a processor and memory.

We can anticipate many potential uses for our system even in a realization closer to its current modest form. Because it draws little power and does not require separate memory storage, our system may be preferable to a CPU- or GPU-simulated neural network when energy or space are at a premium. Furthermore, power consumption is not concentrated (as in a CPU/GPU) but distributed evenly across the learning machine, allowing future versions to massively increase speeds without overheating. Because its function is not encoded in its design, our system may be appropriate for tasks that require on-demand flexibility, for example as a sensor that detects deviations from an as-yet unspecified background signal. Because it is robust to damage, it may be useful for scenarios where the system is exposed to danger. 

Our system is robust to damage because it is composed of many repeated identical elements that update themselves in response to stimuli. It is therefore a kind of ``learning material" or metamaterial in the sense that it is a many-element system with learning as an emergent collective property that is not inherent in the arrangement of its elements nor in the selection of input or output locations. If constructed appropriately, physics-based learning networks should be easily modifiable after construction; just as removing arbitrarily-chosen edges does not destroy functionality, additional edges do not require precise placement to be useful. It is not outlandish to imagine a future adaptive realization of similar learning circuits that would have no need for any \textit{a priori} design in order to learn, and could be augmented or divided like clay. 

\acknowledgements{We thank James MacArthur for advice regarding circuit design and Marc Miskin for instructive discussions, especially regarding scalability. This work was supported by the National Science Foundation via the UPenn MRSEC/DMR-1720530 (S.D. and D.J.D.), DMR-2005749 (A.J.L.) and the U.S. Department of Energy, Office of Basic Energy Sciences, Division of Materials Sciences and Engineering award DE-SC0020963 (M.S.). A provisional patent (No. 63/191,468) has been filed for the design of the physics-driven learning circuit.}

\appendix


\section{Task Details}
\begin{figure} 
\includegraphics[width = 8.7cm]{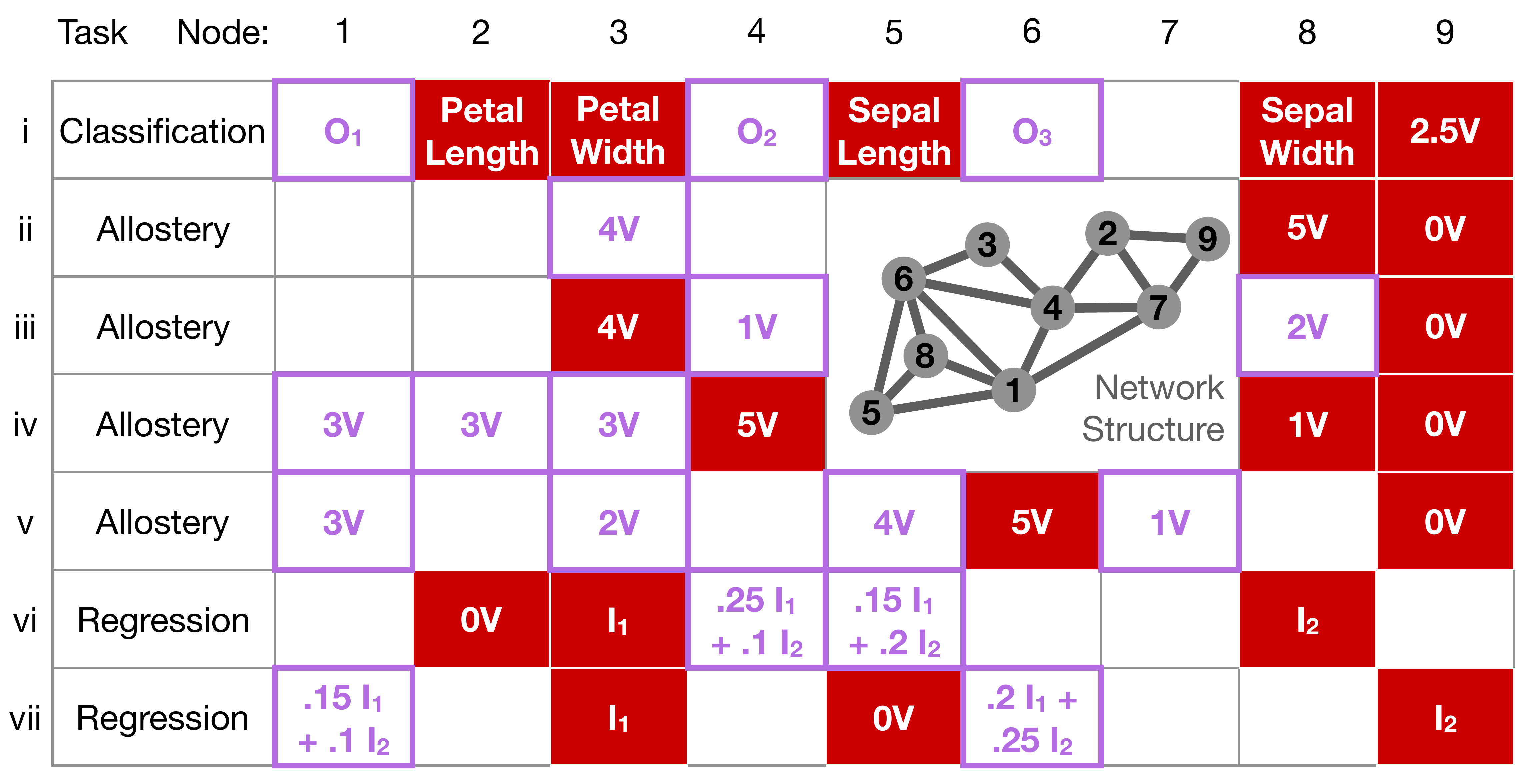}
\caption{\textbf{Task Details.} Voltage specification by node number for each task detailed in Fig. 4. Red cells are input nodes, purple-outlined cells are output nodes. Node numbers correspond to the network structure as shown in the inset. }
\label{figS3}
\end{figure}

Tasks listed in Fig.~4 in the main text are detailed in Fig.~\ref{figS3}. For allosteric tasks, input or desired output voltages are listed as single values. For regression tasks, training and test set inputs are selected using a uniform random distribution between 1 and 5~V, and output desired voltages are functions of these inputs, as listed. For the classification task, each input (e.g. all petal widths) are re-scaled to span 0 to 5~V. A typical classification output scheme in an artificial neural network (ANN) would designate one output node for each class and train towards producing a high value (e.g. 5~V) at the node of the correct class, and 0~s at all other output nodes. However, this output basis is not feasible because our network is linear. We instead choose an output basis as follows: At the start of every epoch (every 30 training steps), we measure the network's output response to the \textit{average} input values from each species of flower in the training set. In a linear network, this is identical to calculating the average output values from all elements in the training set, as done in previous work~\cite{stern_supervised_2020}. During the ensuing epoch, the desired output voltage for each flower is this average response for the appropriate species. These desired voltages evolve as the network trains, but eventually settles at consistent values. Because these output averages depend solely on training data, they may be useful in the future for determining when to stop training a learning network. Furthermore, this averaging method improves the initial accuracy beyond the expected 33\%, since it picks target values with a minimal distance to the network response for a given species.

\section{Scaling the Electronics}
Our prototype was not built with speed or scale as a priority, and as a result leaves much room for improvement in these regards. Our system takes up several square feet, and operates at about 3-5~Hz, limited by the data acquisition and voltage setting hardware. Analog networks utilizing variable weights and comparators (without utilizing physics-as-computation) have been accomplished with under 100 transistors per edge equivalent of our prototype (often referred to as synapses)~\cite{schneider_analog_1993}. State of the art CMOS fabrication can yield roughly 300 million transistors per mm$^2$, operating on nanosecond timescales or faster~\cite{yeap_5nm_2019}. Using these estimates, a 10 million-edge physical learning network could be implemented with a footprint less than 10mm$^2$. Such a system would represent a $10^6$ increase in edge count, a $10^5$ decrease in footprint, and a $10^8$ increase in speed from our prototype.

\begin{figure*} 
\includegraphics[width = 11.4cm]{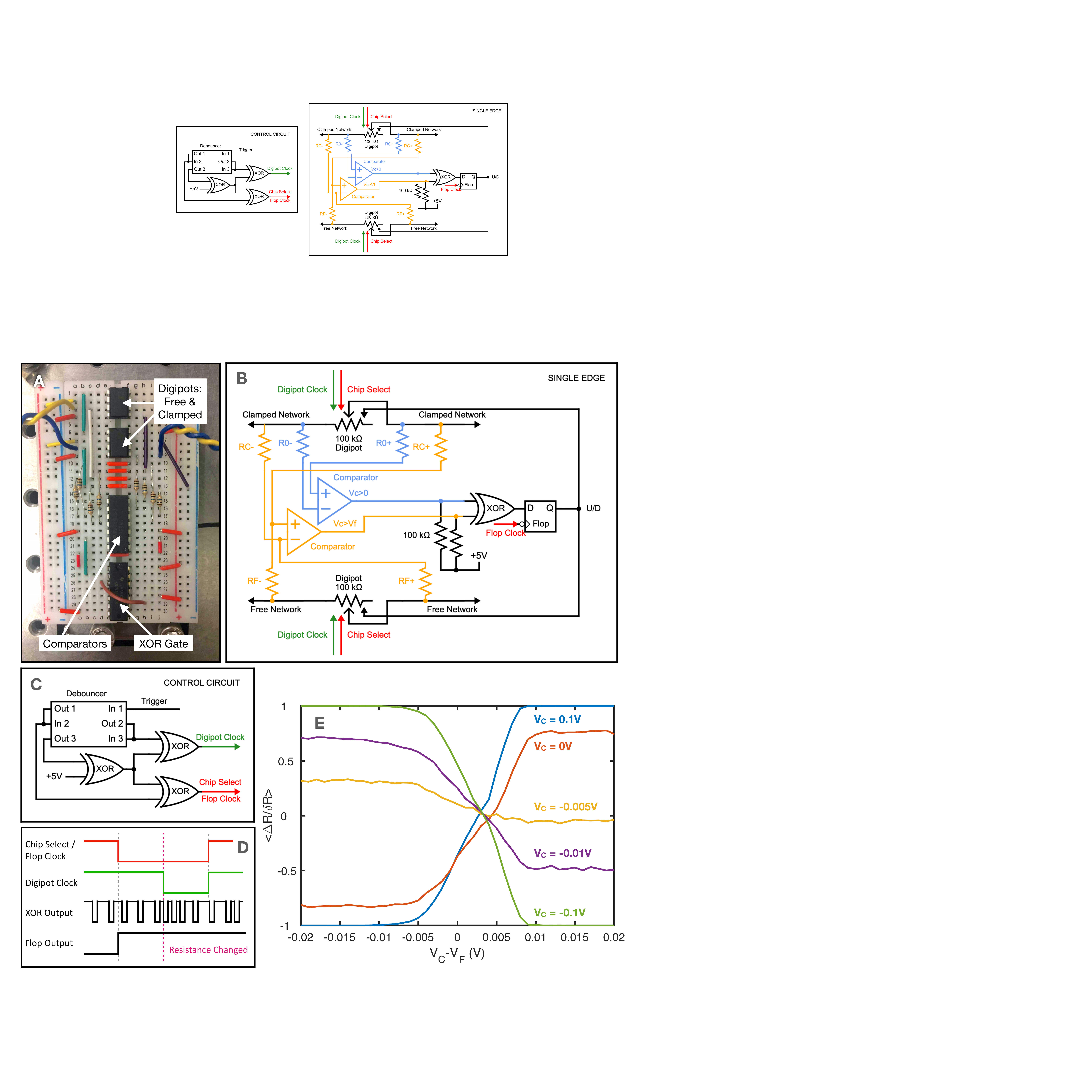}
\caption{
\textbf{A single edge of the network.} 
\textbf{(A)} Image of an edge, as constructed on a breadboard. 
\textbf{(B)} Circuit diagram for a single edge, which houses circuitry for both both the free and clamped networks. Comparators and an XOR gate compute the direction of resistance change based on the relative voltage drops across the free and clamped variable resistors (digipots), and the XOR output is stored in a D-Flop before being fed back into the up/down input of the potentiometers.
\textbf{(C)} Global clock circuitry. The control circuit receives an ascending or descending edge from the data acquisition card (computer) into the 'Trigger' port. This produces a cascading effect through the debouncer, changing output 1, then 2, then 3, which are fed into XOR gates. 
\textbf{(D)} This cascade results in a descending edge in the digipot chip select/D flop clock signal, then a descending edge in the digipot clock signal, and finally a return to high for both signals. As a result, the XOR output of the edge circuit shown in \textbf{B} is sampled and stored by the D~flop ahead of the digipot clock triggering a change in resistance. This avoids feeding the potentially fluctuating XOR signal directly into the digipot. 
\textbf{(E)} Average resistance change as a function of comparator voltages $V_C$ and $(V_C-V_F)$. Ideally we would have step functions jumping at $(V_C-V_F)=0$~V. Noise spreads out the transition, and in this edge, comparator bias shifts the curves to the right.
}\label{figS1}
\end{figure*} 

\section{Circuitry}

Our electrical network uses variable resistors as edges (AD5220 digital potentiometers wired as rheostats). These `digipots' are not continuously adjustable as assumed by the original coupled learning rule~\cite{stern_supervised_2021}, but instead have 128 resistance values evenly spaced by $\delta R = 100K\Omega/128 \sim 781\Omega$. We therefore restrict the evolution of each edge to discrete steps $\pm\delta R$ in either direction. The coupled learning rule then simplifies to
\begin{equation}
    \Delta R_i = 
     \begin{cases}
      +\delta R & \text{if} \ |\Delta V^C_i| > |\Delta V^F_i|, \\
      -\delta R & \text{otherwise.}
    \end{cases}  
    \label{signS}
\end{equation}
Other learning rules that only depend on the signs of the gradient of cost functions have been shown to be successful~\cite{bernstein_signsgd_2018}. This new rule is also easier to implement digitally as it only requires a Boolean comparison of voltage drops instead of a difference in energy dissipation. However, Eq.~(\ref{signS}) still requires access to both the free and clamped electrical states. To this end, we construct two identical networks for comparison, one running the free state and one running the clamped state. Corresponding edges of the free and clamped networks always have the same resistance, and are housed on the same breadboard (Fig.~\ref{figS1}A).

The absolute value comparison in Eq.~(\ref{signS}) is still non-trivial to evaluate electronically. A comparator produces a signed comparison $\Delta V_i^C > \Delta V_i^F$, but this will yield the \textit{opposite} of our desired value if both drops are negative, which we cannot rule out \textit{a priori}. We can, however, assume that the two voltage drops have the same sign. Empirically, we find this is nearly always the case, especially for $\eta \ll 1$. We can then use a second comparison, $\Delta V_i^C < 0$, to determine if $\Delta V_i^C > \Delta V_i^F$ is equivalent to $|\Delta V^C_i| > |\Delta V^F_i|$ (positive voltage) or its inverse (negative voltage). Our learning rule can now be written using only functions of common logical circuit components:
\begin{equation}
     \Delta R_i =   
     \begin{cases}
      +\delta R & \text{if} \ \textrm{XOR}\left[\Delta V_i^C > \Delta V_i^F, 0 < \Delta V_i^C\right] \\
      -\delta R & \text{otherwise}
    \end{cases}  
    \label{simplerule}
\end{equation}
We implement Eq.~(\ref{simplerule}) with two comparators (LM339AN), one XOR gate (SN74ALS86N), and one D-Flop (TI CD74HC73E JK flop plus SN74ALS86N XOR gate) on every edge (Fig.~\ref{figS1}B). On each edge, the output of XOR gate is stored in the D-Flop and fed back into the up/down input of the digital potentiometers in both free and clamped networks. During training, the resistance updates of every variable resistor are triggered by the descending edge of a global clock signal fed into the digital potentiometers. A switch debouncer/delay (MC14490PG) circuit and three XOR gates are wired to generate two sequential descending edge signals (red and green in Fig.~\ref{figS1} B, C, and D). The first descending edge is used to trigger a D flop (TI CD74HC73E JK flop plus SN74ALS86N XOR gate) to store the output of the XOR gate (Eq.~\ref{simplerule}). Because the learning machine naturally moves the voltage of the free and clamped networks towards each other, this XOR output (U/D signal) will typically become dominated by noise by the end of training, and will oscillate rapidly. Storing the value in the D-flop ensures a clean signal in the U/D port of the digital potentiometers, as shown in Fig.~\ref{figS1}B. Finally, the variable resistors (digipots) used in our system (AD5220 100k) have a slight bias in their logical evaluation. As a result, the update rule (Eq.~\ref{simplerule}) is imperfectly evaluated at similar free and clamped voltage drops, as shown in Fig.~\ref{figS1}D. These incorrect evaluations do not prevent our system from functioning, but do limit the error floor.

\bibliography{bib}
\end{document}